\documentclass[twocolumn,amsmath,amssymb,aps,pra,longbibliography,floatfix]{revtex4-1}
\usepackage{physics}
\usepackage{units}
\usepackage{amsmath}
\usepackage{cases}
\usepackage{url}
\usepackage{natbib}
\usepackage{textcase}
\usepackage{amssymb}
\usepackage{graphicx}
\usepackage{bm} 

\usepackage{times}
\usepackage{float}
\usepackage{multirow,microtype,color,relsize}
\usepackage{subfigure}
\usepackage[breaklinks=true]{hyperref}
\usepackage[utf8]{inputenc}
\usepackage[english]{babel}
\hypersetup{colorlinks=true,linkcolor=blue,citecolor=blue}
\hypersetup{linktocpage}
\usepackage{CJKutf8}
\usepackage{breakcites}
\usepackage{float}
\usepackage{siunitx}
\usepackage[dvipsnames]{xcolor}
\definecolor{mypine}{RGB}{1, 121, 111}

\newcommand{\bsts}{BiSbTeSe$_2$~}

\begin{document}
\begin{CJK*}{UTF8}{gbsn}
\title{Disorder effects in topological insulator thin films}

\author{Yi Huang~(黄奕)}

\email[Corresponding author: ]{huan1756@umn.edu}
\author{B.\,I. Shklovskii}
 
\affiliation{School of Physics and Astronomy, University of Minnesota, Minneapolis, Minnesota 55455, USA}
\date{\today}

\begin{abstract}
Thin films of topological insulators (TI) attract large attention because of expected topological effects from the inter-surface hybridization of Dirac points. However, these effects may be depleted by unexpectedly large energy smearing $\Gamma$ of surface Dirac points by the random potential of abundant Coulomb impurities. We show that in a typical TI film with large dielectric constant $\sim 50$ sandwiched between two low dielectric constant layers, the Rytova-Chaplik-Entin-Keldysh modification of the Coulomb potential of a charge impurity allows a larger number of the film impurities to contribute to $\Gamma$. As a result, $\Gamma$ is large and independent of the TI film thickness $d$ for $d > 5$ nm.  In thinner films  $\Gamma$ grows with decreasing $d$ due to reduction of screening by the hybridization gap.  We study the surface conductivity away from the neutrality point and at the neutrality point. In the latter case, we find the maximum TI film thickness at which the hybridization gap is still able to make a TI film insulating and allow observation of the quantum spin Hall effect,  $d_{\max} \sim 7$ nm.
\end{abstract}
\maketitle
\end{CJK*}

\section{Introduction}

Topological insulators (TI) continue to generate a strong interest because of their surfaces host massless Dirac states on the background of the bulk energy gap. 
Typically, as-grown TI crystals are heavily doped semiconductors with concentration of donors $\sim 10^{19}$ cm$^{-3}$. 
(For certainty, we talk about n-type case where the Fermi level is high in the conduction band). 
However, to employ Dirac states in transport, one has to move the Fermi level close to the Dirac point. 
In bulk crystals, this is done by intentional compensation of donors with almost equal concentration of acceptors. 
With increasing degree of compensation, the Fermi level shifts from the conduction band to inside the gap and eventually arrives at the surface Dirac points. 

This seemingly easy solution of the Fermi-level problem, however, comes with a price~\cite{skinner2012}. 
In fully compensated TI, all donors and acceptors are charged, and these charges randomly distributed in space create random potential fluctuations as large as the TI semiconductor gap. 
These fluctuations create equal numbers of electron and hole puddles, and substantially reduce the activation energy of the bulk transport~\cite{ren2011,knispel2017}.
At the same time near the surface, the random potential of charged impurities smears the Dirac point by the energy $\Gamma$ self-consistently determined by the surface electrons screening~\cite{skinner2013a,skinner2013b}. 
This smearing was observed by the scanning tunnel microscopy~\cite{beidenkopf2011}. 
It also should determine the width of Landau levels of Dirac electrons and quantum relaxation time $\tau_q = \hbar/\Gamma$ as measured by Shubnikov-de-Haas oscillations. 

\begin{figure}[t]
    \centering
    \includegraphics[width=\linewidth]{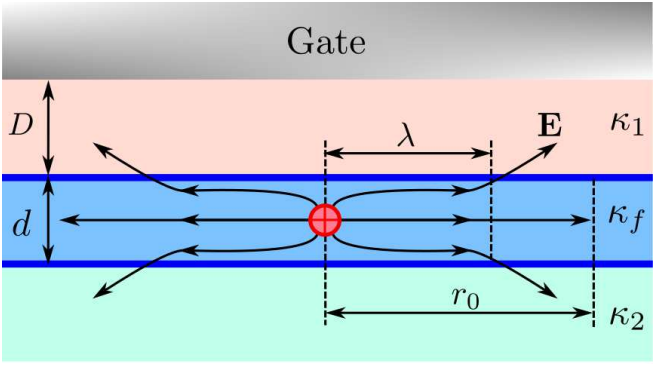}
    \caption{TI thin film of thickness $d$ with dielectric constant $\kappa_f$ deposited on the substrate with dielectric constant $\kappa_2$. The top metallic gate is separated from the film by a spacer of thickness $D$ with dielectric constant $\kappa_1$.  The topological surfaces are shown by blue lines. In the case $\kappa_f \gg \kappa$, a typical charge impurity is shown by a red circle with its electric field $\vb{E}$ (black) channeling through the film for a distance $\lambda$ before exiting outside. In similar topologically trivial films, the electric field exits at a larger distance $r_0$.}
    \label{fig:film}
\end{figure}

Recently TI research shifted to thin TI films of thickness $d < 20$ nm range~\cite{zhang2010,kim2013,nandi2018,chong2020,chaudhuri2020,dibernardo2020,wang2020}. 
This interest is related to observations of the inter-surface hybridization leading to the Dirac points hybridization gaps $\Delta(d)$ and related topological effects, including the quantum spin Hall effect~\cite{chong2020}. However, such observations are obscured by unexpectedly large effects of disorder. 
One might think that the role of disorder in thin TI films should be smaller than in the bulk TI.
Indeed, at a given total 3D concentration of charged impurities $N$, the 2D concentration of them $N d$ in a thin TI film is quite small. 
In a thin film, the Fermi level can be shifted to the Dirac point by the gate parallel to the TI film (see Figure~\ref{fig:film}). 
Therefore, one might expect that the compensation by acceptors can be avoided to get a much smaller $\Gamma(d)$. 
However, a distant gate can only compensate the average charge density of donors. 
Local fluctuations of the donor concentration and charge density still create a large random potential that, after self-consistent screening by surface electrons, results in a large Dirac point smearing energy $\Gamma(d)$. 

In this paper, we show that in a typical TI film with large dielectric constant $\sim 50$ sandwiched between two low dielectric constant layers, the Rytova-Chaplik-Entin-Keldysh modification of the Coulomb potential of a charge impurity slows down the potential decay in space, and allows a larger number of the film impurities to contribute in $\Gamma$. As a result, $\Gamma$ is large and independent on the TI film thickness $d$ for $d > 5$ nm. At smaller thickness, $\Gamma$ grows with decreasing $d$ due to reduced by hybridization gap screening. We also study the surface conductivity both far away from neutrality point when kinetic energy is much larger than $\Gamma$ and at the neutrality point. In the latter case, we find the maximum thickness at which the hybridization gap makes a TI film insulating and allows observation of the quantum spin Hall effect, $d_{\max} \sim 7$ nm.

Contrary to the bulk case, what happens in the thin TI film strongly depends on the average dielectric constant of the film environment $\kappa_{} = (\kappa_1 +\kappa_2)/2$, where indexes 1 and 2 are related to two sides of the film (see Figure~\ref{fig:film}). Below we consider three different cases $\kappa_f \gg \kappa_{}$, $\kappa_f = \kappa_{}$, and $\kappa_f \ll \kappa_{}$. 

In Section II we calculate $\Gamma$ at the neutrality point for the most interesting first case when the potential of a charge impurity was described by Rytova~\cite{rytova1967}, Chaplik and Entin~\cite{chaplik1971}, and Keldysh~\cite{keldysh1979}. In Section III we study the cases $\kappa_f = \kappa_{}$ and $\kappa_f \ll \kappa_{}$. In Section IV we comment on the role of the gate when it is close to the film surface. In Section V we calculate the conductivity of a TI film ignoring hybridization gap. In Section VI we concentrate on the effect of the hybridization gap on the conductivity at the neutrality point and find the maximum thickness $d_{\max}$ at which TI film is still insulating so that one can observe spin Hall effect. 

\section{Thin TI film in small dielectric constant environment}

In this section, we calculate $\Gamma(d)$ in the case of $\kappa_f \gg \kappa_{}$. 
For example~\cite{wang2020,chong2020}, a \bsts (BSTS) thin film with $\kappa_f \sim 50$ can be sandwiched between two h-BN layers with $\kappa_{1,2} \sim 5$~\cite{laturia2018}.
In this case $\kappa_{} \sim 5$ is 10 times smaller than $\kappa_f$.
If $\kappa_f \gg \kappa_{}$, the electric field of a charged impurity inside the thin film is trapped inside the film for a distance $r_0 = (\kappa_f/2\kappa_{})d$, and only after $r > r_0$ the electric field exits to the environment. 
This leads to the effective Coulomb interaction with asymptotic expressions~\cite{chaplik1971},
\begin{equation}\label{eq:keldysh}
    v_0(\vb{r}) \approx 
    \begin{cases}
    \frac{e^2}{\kappa_{} r}\qc &r> r_0,\\
    -\frac{e^2}{\kappa_{} r_0}\qty[\ln(r/2r_0) + \gamma]\qc &d < r < r_0,
    \end{cases}
\end{equation}
where $\vb{r}$ is a 2D vector in the plane of TI film, and $\gamma = 0.577$ is the Euler constant.
The Fourier transform of $v_0(\vb{r})$ is 
\begin{equation}\label{eq:fourier_keldysh}
    v_0(q) = \frac{2\pi e^2}{\kappa_{} q (1+ q r_0)},
\end{equation}
valid for $q < 1 /d$.

In a TI film, the electric field of a charged impurity experiences additional screening by  Dirac electrons living on the surfaces of the film.
To describe this screening, we start from the equation for the electric potential of screened charged impurities $\phi(\vb{r})$
\begin{equation}
   \mu [n(\vb{r})] - e \phi(\vb{r}) = E_F,
\end{equation}
where $E_F={\rm const.}$ is the Fermi level (electro-chemical potential), $\mu[n(\vb{r})] =\hbar v_F k_F[n(\vb{r})]$ is the (local) chemical potential, $v_F$ is the velocity near the Dirac cone, and $k_F[n(\vb{r})] = \sqrt{4 \pi \abs{n(\vb{r})}}$ is the local Fermi wave vector.
If the average chemical potential $\mu$ is large enough, so that $\mu^2 \gg e^2\phi^2$, then $\mu[n(\vb{r})]$ can be linearized in the local carrier density variation $\delta n (\vb{r})$
\begin{align}
    \mu[n(\vb{r})] \approx \mu + \delta n(\vb{r})/\nu(\mu).
\end{align}
where $\nu(\mu) = d{n}/d{\mu}=  \mu / (2\pi (\hbar v_F)^2)$ is the thermodynamic density of states (TDOS) at zero temperature.
Introducing the effective fine structure constant $\alpha = e^2 / \kappa_f \hbar v_F$, we can write the TDOS as 
\begin{equation}\label{eq:tdos}
    \nu(\mu) = \frac{\kappa_f^2 \alpha^2}{2\pi e^4} \mu. 
\end{equation}
In the Thomas-Fermi (TF) approximation~\footnote{TF approximation is justified as long as $\alpha \ll 1$. See related discussion in Ref.~\onlinecite{skinner2013a}.}, the screening by surface electrons can be described by the dielectric function
\begin{equation}
    \epsilon(q) = 1- v_0(q) \Pi_{TF},
\end{equation}
where the TF polarization bubble is $\Pi_{TF} = -\nu(\mu)$, and the bare interaction $v_0(q)$ is given by Eq.~\eqref{eq:fourier_keldysh}.
We arrive at the screened potential of one charge impurity within a thin TI film
\begin{equation}\label{eq:vq}
    v(q) = \frac{v_0(q)}{\epsilon(q)} = \frac{2\pi e^2}{\kappa_{} [q (1+ q r_0) + q_s]},
\end{equation}
where $q_s = 2\pi e^2 \nu/\kappa $ and $q < 1 /d$.

We see that if $q_s r_0 \gg 1$ then, unlike in uniform 3D dielectrics, inside the TI film a strong screening happens at the distance 
\begin{equation}
    \lambda = (r_0 /q_s)^{1/2}.
\end{equation}
Indeed, the behavior of $v(q)$ changes at $q = \lambda^{-1}$:
\begin{equation}
    v(q) \approx 
    \begin{cases}
    \frac{2\pi e^2 }{\kappa_{} q_s}\qc & q< \lambda^{-1}, \\
    \frac{2\pi e^2}{\kappa_{} q^2 r_0}\qc & \lambda^{-1}< q < d^{-1}.
    \end{cases}
\end{equation}

\begin{figure}[t]
    \centering
    \includegraphics[width = \linewidth]{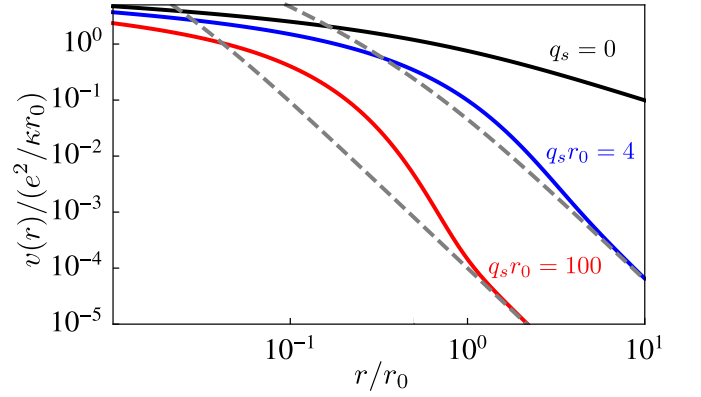}
    \caption{Log-log plot of the screened interaction $v(r)$ for different $q_s$. The gray lines are obtained by the Fourier transform of $2\pi e^2 / \kappa_{}(q+q_s)$, which shows $\sim r^{-3}$ in large distance.}
    \label{fig:screening}
\end{figure}
The behavior of $v(r)$ for different values of $q_s$ is shown in Figure~\ref{fig:screening}. At large distance $r \gg \lambda$, we get $v(r) \simeq e^2 /\kappa_{} q_s^2 r^3$ like for a quadrupole. The difference between topological and topologically trivial films is also schematically illustrated in Figure~\ref{fig:film}.

Assuming that impurities are randomly distributed inside the film, the mean squared fluctuation of the potential is given
by~\footnote{Here we drop the short distance contribution $r < d$ to the potential $\phi(\vb{r})$ which is standard 
Coulomb potential. This does not change the result substantially as long as $d \ll \lambda$.}
\begin{align}
    \ev{\phi^2} &= \frac{1}{e^2}\int  d\vb{r} v^2(\vb{r}) Nd  \nonumber \\
    &= \frac{2\pi N d e^2}{\kappa_{}^2} f(q_s r_0),
\end{align}
where the function $f(x)$ reads
\begin{align}
    f(x) &= \frac{2}{4x - 1} \nonumber\\
    &+
    \begin{cases}
    \frac{2}{(1-4x)^{3/2}}\tanh^{-1}{\sqrt{1-4x}}\qc & 0<x<1/4,\\
    -\frac{2}{(4x-1)^{3/2}}\tan^{-1}{\sqrt{4x-1}}\qc & x>1/4.
    \end{cases}
\end{align}
We are interested in two limiting cases of the dimensionless parameter $q_s r_0 = (r_0/\lambda)^2$:
\begin{equation}\label{eq:phi2limit}
    \ev{\phi^2} = \frac{2\pi N d e^2}{\kappa_{}^2}
    \begin{cases}
    (2 q_s r_0)^{-1}\qc &  \lambda \ll r_0, \\
    -2-\ln(q_s r_0)\qc &  \lambda \gg r_0.
    \end{cases}
\end{equation}
There is a simple qualitative interpretation of the limiting expression of $\ev{\phi^2}$.
In the case when $\lambda \ll r_0$ (or $q_s r_0 \gg 1$), surface electrons screening cuts off the impurity potential at distance $\lambda$ from the impurity center.
The fluctuation of number of impurities inside radius $\lambda$ is equal to $ (N d \lambda^2)^{1/2}$.
Since each charge impurity of this area contributes to the potential $\sim e/\kappa_{} r_0$ [see Eq.~\eqref{eq:keldysh}], we get $\ev{\phi^2} \sim (N d \lambda^2) (e/\kappa_{} r_0)^2$, namely the first line of Eq.~\eqref{eq:phi2limit}. 
On the other hand, at $\lambda \gg r_0$ (or $q_s r_0 \ll 1$) the potential of impurity  $v(\vb{r})$ follows Eq.~\eqref{eq:keldysh} with effective screening length $r_0$. Taking into account that fluctuation of number of impurities inside radius $r_0$ is $\sim \sqrt{N d r_0^2}$, we arrive at the second line of Eq.~\eqref{eq:phi2limit}.

We are interested in the charge neutrality point where $E_F = 0$, and $\phi$ has the Gaussian distribution function with $\ev{\phi} = 0$ and $\ev{\phi^2} = \Gamma^2/e^2$. 
Next, we want to calculate the average density of states $\ev{\nu}$ using the Gaussian distribution function of $\phi$,
\begin{equation}
    \ev{\nu} = \int_{-\infty}^{\infty} d(e\phi) 2\nu(e\phi) \frac{e^{-e^2\phi^2/2\Gamma^2}}{\sqrt{2\pi \Gamma^2}} = \frac{2 \alpha^2 \kappa_f^2 \Gamma}{(2\pi^3)^{1/2} e^4}.
\end{equation}
Here we multiply the density of states by a factor of 2 because the potential at each surface is screened by electrons of both the top and bottom TI surfaces. 
The above use of the potential $\sim e^2/\kappa r_0$ inside the the TI film at distance $d < r < \lambda$ from a Coulomb impurity apparently is valid only for $d < \lambda$. This condition is equivalent to $d \lesssim d_c = \alpha^{-4/3}N^{-1/3}$~\footnote{The value of $d_c$ will be determined after we obtained $\lambda(d)$ self-consistently in Eq.~\eqref{eq:lambdaneutral}}. 
For thicker films, $d > d_c$, one should think about two separate surfaces like in a bulk sample where each surface screens its own random potential~\cite{skinner2013a}. 
Then one also can find the lower limit of applicability of large $d$ theory~\cite{skinner2013a}, $d_c$, as the screening radius $r_s$ of a single surface found in Ref.~\cite{skinner2013a}.

At $d \lesssim d_c$ replacing $\nu$ by $\ev{\nu}$ in $q_s = 2\pi e^2 \nu/\kappa $, we have
\begin{equation}\label{eq:qs}
    q_s = \sqrt{\frac{8}{\pi}} \frac{ \alpha^2 \kappa_f^2 \Gamma}{\kappa_{} e^2}.
\end{equation}
Now one can solve for $\Gamma$ and $q_s$ self-consistently using Eqs.~\eqref{eq:phi2limit} and~\eqref{eq:qs}.
If $\lambda \ll r_0$, then 
\begin{align}\label{eq:gamma_gg1}
    \Gamma &= \qty(\frac{\pi^3}{2})^{1/6} \frac{e^2 N^{1/3}}{\kappa_f \alpha^{2/3}}, \\
    q_s &= 2^{4/3} \alpha^{4/3} \frac{\kappa_f}{\kappa_{}} N^{1/3},\label{eq:qs_gg1} \\
    \lambda &= 2^{-7/6} \alpha^{-2/3} (Nd^3)^{-1/6} d. \label{eq:lambdaneutral}
\end{align}
The result for $\Gamma$ is independent on $d$, and up to a numerical factor is the same as the results earlier obtained for a bulk samples~\cite{skinner2013a}. 
Therefore, our $\Gamma$  easily matches that of Ref.~\cite{skinner2013a} at $d=d_c$. To ensure the self-consistency, one should check whether the assumption $q_s r_0 \gg 1$ with $q_s$ given by Eq.~\eqref{eq:qs_gg1} is correct. We find that, Eqs.~\eqref{eq:gamma_gg1}, \eqref{eq:qs_gg1} and \eqref{eq:lambdaneutral} are valid if $d \gg d_1 = (\kappa_{}/\kappa_f)^2 \alpha^{-4/3} N^{-1/3}$.
Note that at the neutrality point Eq. \eqref{eq:lambdaneutral} provides the typical size of puddles, while the concentration of electrons and holes in puddles is $n_p \sim (Nd\lambda^2)^{1/2}/\lambda^2 \sim (\alpha N)^{2/3}$. This concentration does not depend on $d$ and is the same as the puddle concentration at the surface of a bulk sample~\cite{skinner2013a}.  

In the other limiting case $\lambda \gg r_0$, in the first approximation  we have
\begin{align}\label{eq:gamma_ll1}
    \Gamma &\approx \qty{\frac{2\pi Nd e^4}{\kappa_{}^2} \ln\qty[\qty(\frac{\kappa_{}}{\kappa_f})^3 \frac{1}{2 \alpha^2 (Nd^3)^{1/2}}]}^{1/2}, \\
    q_s &\approx 4 \qty(\frac{\alpha \kappa _f}{\kappa_{}})^2 \sqrt{Nd} \qty{\ln\qty[\qty(\frac{\kappa_{}}{\kappa_f})^3 \frac{1}{2\alpha^2 (Nd^3)^{1/2}}]}^{1/2}.\label{eq:qs_ll1}
\end{align}
Eqs.~\eqref{eq:gamma_ll1} and \eqref{eq:qs_ll1} are valid if $d \ll d_1$, i.e., the arguments of logarithms are much larger than unity.

\begin{figure}[t]
    \centering
    \includegraphics[width = \linewidth]{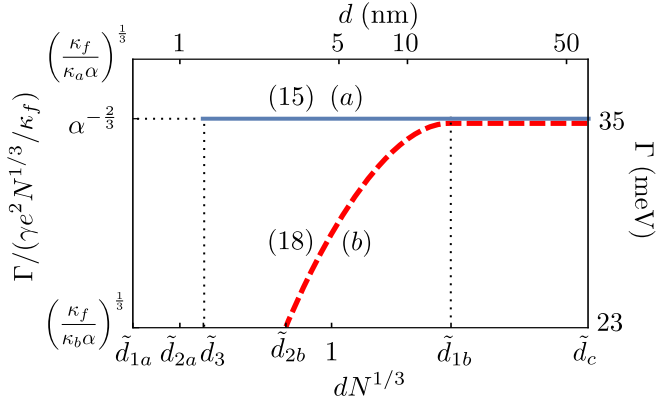}
    \caption{Log-log plot of the disorder potential amplitude $\Gamma$ as a function of thickness $d$ for two examples of scenarios (a) and (b). We use $N = 10^{19}$ cm$^{-3}$, $\kappa_f = 50$, and $\alpha^{-1} = 7$ in order to relate the dimensionless left-vertical and bottom-horizontal axes to the right-vertical and top-horizontal dimensional axes. The blue solid line corresponds to the scenario (a): $\kappa_f/\kappa_{} > \alpha^{-1} > 1$ (in this example we use such $\kappa = \kappa_a$ that $\kappa_f/\kappa_{a} = 10$). The red dashed curve corresponds to the alternative scenario (b): $\alpha^{-1} > \kappa_f/\kappa_{} \geq 1$ (for this example we use such $\kappa = \kappa_b$ that $\kappa_f/\kappa_{b} = 2$ ). 
    On the left-vertical axis, dimensionless potential amplitude $\Gamma$ in units of $\gamma e^2 N^{1/3}/\kappa_f$ where $\gamma = (\pi^3/2)^{1/6}$ is shown.
    On the right-vertical axis, we show $\Gamma$ in units of meV.
    On the bottom-horizontal axis we show the characteristic dimensionless TI film widths $\tilde d_{1a,b} = (\kappa_{a,b}/\kappa_f)^2 \alpha^{-4/3}$, $\tilde d_{2a,b} = (\kappa_{a,b}/\kappa_f)^{2/3}$, $\tilde d_3 = \alpha^{2/3}$, and $\tilde d_c= \alpha^{-4/3}$. 
    On the top-horizontal axis, the film thickness $d$ in units of nm is shown.
    Numbers of Eqs.~\eqref{eq:gamma_gg1} and~\eqref{eq:gamma_ll1} describing parts of $\Gamma(d)$ plots are shown next to them. }
    \label{fig:gamma_d}
\end{figure}

In order to derive the above results we assumed that electric potential fluctuations follow Gaussian distribution.
This assumption is valid if the number of substantially contributing to the potential impurities $M \gg 1$.
If $\lambda > r_0$, or $q_s r_0 <1$, then Eq.~\eqref{eq:qs_ll1} yields $M=N d r_0^2 \sim N d^3 (\kappa_f/\kappa_{})^2 \gg 1$ when $d \gg d_2 = (\kappa_{} / \kappa_f)^{2/3} N^{-1/3}$.
On the other hand, if $\lambda < r_0$, or $q_s r_0 >1$, using Eq.~\eqref{eq:lambdaneutral} we get that $M = Nd \lambda^2 \sim (N d^3)^{2/3} \alpha^{-4/3} \gg 1$ when $d \gg d_3 = \alpha^{2/3} N^{-1/3}$.
 
In Figure~\ref{fig:gamma_d} we schematically summarize our results for $\Gamma(d)$ for two scenarios covering generic situation with $\alpha^{-1} > 1$ and $\kappa_f/\kappa_{} >1$:

Scenario (a) is defined by inequality $\kappa_f/\kappa_{} > \alpha^{-1} > 1$. In this scenario the energy $\Gamma(d)$ is a constant given by Eq.~\eqref{eq:gamma_gg1} for $d > \alpha^{2/3} N^{-1/3}$, while for $d < \alpha^{2/3} N^{-1/3}$ the Gaussian approach fails. 

Scenario (b) is defined by inequality $ \alpha^{-1}>\kappa_f/\kappa_{} \geq 1$. In this case the energy $\Gamma(d)$ is a constant given by Eq.~\eqref{eq:gamma_gg1} for $d > d_1 = (\kappa_{}/\kappa_f)^2 \alpha^{-4/3} N^{-1/3}$, while  $\Gamma(d)$ crosses over to Eq.~\eqref{eq:gamma_ll1} for $(\kappa_{}/\kappa_f)^{2/3} N^{-1/3} =d_2 < d < d_1= (\kappa_{}/\kappa_f)^2 \alpha^{-4/3} N^{-1/3}$. 
In this scenario, the Gaussian approach fails at $d < d_2$. Scenario (b) includes the case $\kappa_{} = \kappa_f$ for which $d_{2}=N^{-1/3}$ and $d_{1}=d_{c}$, so that for thin films, $d< d_c$,  $d$-independent part of the function $\Gamma(d)$ does not exist.

Let us see how these two scenarios work for TIs based on BSTS-like systems with $v_F \sim 3 \times 10^{5}$ m/s, $\kappa_f \sim 50$, $\alpha^{-1} \sim 7$ and $N \simeq 10^{19}$ cm$^{-3}$. 
If such a TI film is sliced between h-BN layers, then $\kappa \simeq \kappa_a = 5$~\cite{laturia2018}, $\kappa_f/\kappa \simeq 10 > \alpha^{-1}$ bringing us to scenario (a). 
If the same film is sliced between two layers of HfO$_2$ with $\kappa \simeq \kappa_b=25$~\cite{robertson2004}, then $\kappa_f/\kappa \simeq 2 < \alpha^{-1}$ and we find ourselves in scenario (b). These two examples are used in Figure~\ref{fig:gamma_d} to plot functions $\Gamma(d)$ for both scenarios. 
In both scenarios, $d_c \sim \alpha^{-4/3} N^{-1/3} \gg d_{1,2,3}$ is the largest length scale.

In the first example, Eq.~\eqref{eq:gamma_gg1} obtained for bulk samples~\cite{skinner2013a} gives $\Gamma \sim 35$ meV which remains valid till very small film widths $d \sim d_3 \simeq 2$ nm, in spite of smaller concentration of impurities $Nd$. Such an unexpectedly strong role of disorder in thin  BSTS-like TI films sandwiched between two low-$\kappa$ layers is a result of the dielectric constant contrast between the TI film and its environment leading to the large contribution from distant Coulomb impurities into potential fluctuations.

In the above we ignored the concentration of charged impurities in the environment outside the TI film $N_e$. Let us now evaluate the role of such impurities. 
To save electrostatic energy, the electric field lines of an impurity at distance $z \lesssim r_0$ from the film surface first enter inside the TI film and then radially spread inside the film to distance $\sim r_0$ before exiting outside the film to infinity. 
Thus, one can think that effectively each outside impurity is represented inside the film by a charge $e$ disk, with radius $z$ and thickness $d$. 
In the presence of screening, only small minority of the outside impurities with $z < \lambda$ contribute in fluctuating charge of the volume $d\lambda^2$. 
As a result, total effective concentration of impurities projected from outside the film is $N_e\lambda/d$. 
If $N_e\lambda/d< N$, where $\lambda$ is given by Eq.~\eqref{eq:lambdaneutral}, outside impurities can be ignored and our results are valid.
For example, in our scenario (a), for the BSTS TI film with $d \sim 5$ nm on silicon oxide substrate with~\cite{shklovskii2007} $N_e \sim 10^{17}$ cm$^{-3}$, $\lambda/d\sim 2$ and our results are valid~\footnote{On the other hand, if $N_e\lambda/d> N$ the screening length $\lambda$ should be recalculated self-consistently together with $\Gamma$. Then, instead of Eqs.~\eqref{eq:gamma_gg1} and \eqref{eq:lambdaneutral}, we arrive at new results $\Gamma \sim (e^2 N_e^{1/3} / \kappa_f) \alpha^{-6/7} (N_e d^3)^{-1/21}$ and $\lambda \sim d \alpha^{-4/7} (N_e d^3)^{-1/7}$.}. 

Let us now discuss the effect of the hybridization gap on the disorder potential.
In a thin enough clean TI film the surface states of two opposite surfaces hybridize and their Dirac spectra acquires the hybridization gaps 
\begin{equation}\label{eq:delta}
\Delta(d) = \Delta_0 \exp(-d/d_0),  
\end{equation}
where $\Delta_0 \sim 1$ eV and $d_0\sim 1.8$ nm for Bi$_{0.7}$Sb$_{1.3}$Te$_{1.05}$Se$_{1.95}$ while $d_0 \sim 1.2$ nm for BiSbTe$_{1.5}$Se$_{1.5}$~\cite{chong2020}. Such a hybridization gaps $\Delta(d)$ equals to the scenario (a) $\Gamma(d)=35$ meV at $d=4$ nm for $d_0=1.2$ nm, or $d=6$ nm for $d_0 = 1.8$ nm. Thus, our results should be valid for  $d \gtrsim 5$ nm. 
 
More accurately speaking in the presence of gap, the local dispersion relation at the Fermi level becomes $\mu(\vb{r}) = \sqrt{\hbar^2 v_F^2 k_F^2(\vb{r}) + \Delta^2/4}$, and the corresponding thermodynamic density of states is given by Eq.~\eqref{eq:tdos} multiplied by a Heaviside theta function $\Theta(\mu - \Delta/2)$. Below we use $\Gamma_{\Delta}$ for disorder potential amplitude in presence of hybridization gap $\Delta$ while we continue to use $\Gamma(d)$ for obtained above result for $\Delta=0$. If $\Delta \ll \Gamma(d)$, or $d \gg 5$ nm, one gets perturbatively~\cite{skinner2013b} $\Gamma_{\Delta} \approx \Gamma(d) [1 + \Delta^2/24 \Gamma^2(d)]$.
On the other hand, if $\Delta \gg \Gamma(d)$, or $d \ll 5$ nm, surface electrons and holes screen the disorder potential in a nonlinear way, only when it exceeds $\Delta/2$. Such nonlinear screening leads to $\Gamma_{\Delta} \simeq \Delta/2 > \Gamma(d)$.


\section{Thin TI film in the same or larger dielectric-constant environment}

In this section, we first consider the case when $\kappa_f=\kappa_{ }$ and the Coulomb interaction with a charged impurity is $v_0(r) = e^2/\kappa r$.
In the TF approximation, the interaction screened by the surface electrons is given by
\begin{equation}
    v(r, z) = \frac{e^2}{\kappa } \int_0^{\infty} dq \frac{J_0(qr)}{1+q_s/q} e^{-qz},
\end{equation}
where $J_0(x)$ is the zeroth Bessel function of the first kind.
The potential fluctuation squared reads
\begin{align}\label{eq:phi2_uniform}
    \ev{\phi^2} &= \frac{1}{e^2}\int N d\vb{r} \int_0^{d} dz v^2(r,z) \nonumber \\
    &= \frac{2\pi Nd e^2}{\kappa^2} [-e^{2q_s d} {\rm Ei} (-2q_s d)],
\end{align}
where ${\rm Ei}(x)$ is the exponential integral function.
Eq.~\eqref{eq:phi2_uniform} has the following limits
\begin{equation}
    \ev{\phi^2} = 
    \frac{2\pi Nd e^2}{\kappa^2}
    \begin{cases}
    (2q_s d)^{-1}\qc & q_s d \gg 1, \\
    -\gamma - \ln(2q_s d)\qc & q_s d \ll 1.
    \end{cases}
\end{equation}
Next, we solve $\Gamma$ and $q_s$ self-consistently similarly to previous sections.
If $q_s d \gg 1$ one obtains the results for $\Gamma$ and $q_s$ given by Eqs.~\eqref{eq:gamma_gg1} and \eqref{eq:qs_gg1} with $\kappa_f = \kappa_{}$ and smaller by $2^{1/3}$ in coefficients. 
On the other hand, if $q_s d \ll 1$ one gets the results of $\Gamma$ and $q_s$ given by
Eqs.~\eqref{eq:gamma_ll1} and \eqref{eq:qs_ll1}, with $\kappa_f = \kappa_{}$. 
The first solution exists only at $d>d_c$, i.e. in the bulk case, while the second corresponds to the thin film case, $d<d_c$.
Thus, for  $\kappa_f=\kappa_{ }$ we arrived to the same result as in Section II, scenario (b).

For BSTS film with $\kappa_f \sim 50$, $\alpha^{-1} \sim 7$ and the impurities concentration $N \sim 10^{19}$ cm$^{-3}$ surrounded by the dielectrics with $\kappa \sim \kappa_f$, we get $\Gamma \sim 30$ meV at $d > q_s^{-1} \sim 20$ nm.

Let us now briefly consider the case of large-$\kappa$ environment, when $\kappa_f \ll \kappa_{1},\kappa_{2}$.
For example, we can imagine thin TI films sandwiched between two STO layers which have very large dielectric constant. 
They should screen the random potential of impurities $\Gamma_1$ and $\Gamma_2$ on both side 1 and 2 surfaces and make $\Gamma_{1,2} \ll e^{2}N^{1/3}/\kappa_f$.

If STO is only on side 2 of the TI film, it dramatically reduces $\Gamma_2$ of this side, while on the other side potential is screened by STO only at the distance $r > d$. 
The number of impurities contributing to $\Gamma_{1}(d)$ is $\sim \sqrt{Nd^3}$, so that $\Gamma_{1}(d)\sim (e^2/\kappa_{f} d) \sqrt{Nd^3}$.

For BSTS film with $\kappa_f \sim 50$, $\alpha^{-1} \sim 7$, and impurities concentration $N \sim 10^{19}$ cm$^{-3}$ sitting on top of STO, we have $\Gamma_1(d) \sim 3\sqrt{d}$ meV where $d$ is measured in units of nm. For example, if $d = 10$ nm, then $\Gamma_1 \sim 10$ meV.

\section{metallic gate}

In this section, we return to the case $\kappa_f \gg \kappa_{1,2}$ and discuss the effect on $\Gamma$ from the metallic gate on top of the low dielectric constant layer with thickness $D$ (see Figure~\ref{fig:film}). 
To get some intuitions, we will start from the question how such a gate affects the electric field of a point charge inside the film, namely, how the gate modifies the Rytova-Keldysh potential Eq.~\eqref{eq:fourier_keldysh} for the case of topologically trivial semiconductor film without surface electrons and their screening. 
This question was carefully studied in Ref.~\cite{kondovych2017}. 
The main result is that, at small enough separation $D < 4d \kappa_f \kappa_1/ \kappa_2^2 \sim r_0$, large distance part of the potential Eq.~\eqref{eq:fourier_keldysh} is truncated (screened) at the distance $\Lambda=\sqrt{D d\kappa_f/\kappa_1} \lesssim r_0$. 
This happens because electric field lines exits the film in the direction to the gate at the distance $r \gtrsim \Lambda$~\footnote{The length $\Lambda$ can be found also via the following simple variational estimate. 
Let us assume that electric field lines are confined inside the TI film within radius $\Lambda \ll r_0 $ from the point charge 
and exit from the film to the gate in the area $\sim \pi \Lambda^{2}$ . The total electrostatic energy consists of two major contributions, one is the energy of the field inside the film $\sim (e^2/\kappa_f d) \ln(\Lambda/d)$, and the other one is the field energy in the gate dielectric $\sim e^2 D / \kappa_1 \Lambda^2$. Minimizing the total energy yields $\Lambda = \sqrt{D d\kappa_f/\kappa_1} \sim \sqrt{r_0 D}$.}.

Let us now recall what surface electrons screening does to the point charge potential in a TI film without gate. 
We saw in Section II that TI surface electrons screening length is given by $\lambda =\sqrt{r_0/q_s}$. 
Now for a TI film with the gate we have both gate and surface electron screening working together. 
Comparing the expressions of $\Lambda$ and $\lambda$ we see, as one could expect, that the distance $D (2\kappa/\kappa_1) \sim D$, which is essentially the distance to the gate, should play the role of $q_s^{-1}$~\footnote{The factor $(2\kappa/\kappa_1)$ is of order unity if $\kappa_1 \simeq \kappa_2$ are not quite different.}. 
This means that if $D \gtrsim q_s^{-1}$ the gate plays only a perturbative role, while for in the case $D \lesssim q_s^{-1}$ the distance $D$ should replace $q_s^{-1}$ in the final result for $\Gamma$. 
Replacing $q_s^{-1}$ by $D (2\kappa/\kappa_1)$ in the case of $\lambda \ll r_0$ in Eq.~\eqref{eq:phi2limit} yields
\begin{equation}
    \Gamma =2 \qty(\frac{\pi e^4 N D}{\kappa_f \kappa_1})^{1/2}.
\end{equation}
This result is valid when it is smaller than Eq.~\eqref{eq:gamma_gg1}, i.e. at $ D \lesssim D_c = N^{-1/3}\alpha^{-4/3}(\kappa_1/\kappa_f)$.

In most experiments, $D > D_c$, so the screening by gate is negligible compared to surface electrons screening. 
For example, in Ref.~\cite{chong2020}, the gate separation is $D \simeq 20$ nm, while $D_c \sim 5$ nm assuming $\kappa_f/\kappa_1 \sim 10$ and $N \sim 10^{19}$ cm$^{-3}$.

\section{conductivity}
In this section, we calculate the conductivity of the surface for the scenario (a) in section II assuming that $D > D_c$.
In the linear screening region $\mu^2 \gg e^2\phi^2$, where the electron density is weakly perturbed by impurities, using Boltzmann kinetic equation for Dirac electrons, one has the expression of the conductivity for a single surface~\cite{culcer2008},
\begin{equation}\label{eq:conduct}
    \sigma = \frac{e^2}{h} \frac{\mu \tau}{4\hbar}.
\end{equation}
Here $\tau$ is the transport relaxation time whose inverse is given by
\begin{equation}
    \frac{1}{\tau} = \frac{\alpha \kappa_f Nd k_F}{\pi \hbar e^2} \int_0^{\pi} d\theta v^2(q) (1-\cos \theta)\frac{1}{2}(1+\cos \theta), \label{eq:rate}
\end{equation}
where $v(q)$ is given by Eq.~\eqref{eq:vq} with $q = 2k_F \sin \theta/2$ and $q_s = k_F \alpha \kappa_f/\kappa$.
The factor $(1 + \cos \theta)/2$ in Eq.~\eqref{eq:rate} arises when the backscattering is suppressed as a consequence of the spin texture at the Dirac point, as in Weyl semimetals~\cite{burkov2011}. 
Changing the integral variable from $\theta$ to $q$, Eq.~\eqref{eq:rate} can be rewritten as 
\begin{equation}\label{eq:rate2}
    \frac{1}{\tau} = \frac{4\pi e^2\kappa_f \alpha Nd}{\kappa^2 \hbar k_F}  \int_0^{2k_F} \frac{dq}{2k_F} \frac{q^2\sqrt{1-(q/2k_F)^2}}{\qty[q(1+qr_0) + q_s]^2}
\end{equation}
Using $x = q/2k_F$, the integral in Eq.~\eqref{eq:rate2} can be expressed in a dimensionless form
\begin{equation}\label{eq:integral}
    I = \int_0^1 dx\frac{ x^2 \sqrt{1-x^2}}{[x(1+2k_Fr_0x) + (\alpha r_0/d)]^2}.
\end{equation}
In the scenario (a) we are considering $d > d_3 = \alpha^{2/3} N^{-1/3}$ and $d < d_c = \alpha^{-4/3}N^{-1/3}$.
First inequality means that $k_F d > \alpha$, because $k_F > \Gamma/\hbar v_F \simeq (\alpha N)^{1/3}$.
Therefore we are interested in Eq.~\eqref{eq:integral} in the limit $k_F d \gg \alpha$. 
In this case $k_F r_0 \gg \alpha \kappa_f/\kappa >1$, so the integral in Eq.~\eqref{eq:integral} is approximated by
\begin{equation}
    I \approx \int_0^1 dx\frac{ x^2 \sqrt{1-x^2}}{[2k_Fr_0x^2 + (\alpha r_0/d)]^2}
\end{equation}
The integral kernel peaks at $x \simeq (\alpha/2k_F d)^{1/2} \simeq (2k_F \lambda)^{-1}$, which corresponds to a momentum transfer $q_{\max} \simeq \lambda^{-1} \ll k_F$. [Here we used Eq.~\eqref{eq:lambdaneutral} for $\lambda$]. The peak value is $(4q_s r_0)^{-1} = (2k_F d \alpha \kappa_f^2/\kappa^2)^{-1}$.
The width of the peak is $\Delta x \sim (k_F \lambda)^{-1}$.
As a result, the integral in limit $k_F d \gg \alpha$ is given by 
\begin{equation}
    I \approx \frac{\pi}{2 \sqrt{2}} \frac{\kappa^2/\kappa_f^2}{(k_F d)^{3/2} \alpha^{1/2}}.
\end{equation}
and Eq.~\eqref{eq:rate2} is 
\begin{equation}\label{eq:rate3}
    \frac{1}{\tau} \approx \sqrt{2}\pi^2 \frac{\alpha^{1/2} e^2 N}{\kappa_f \hbar k_F^{5/2} d^{1/2}}.
\end{equation}
Substituting Eq.~\eqref{eq:rate3} into Eq.~\eqref{eq:conduct} with $k_F = \sqrt{4\pi n}$, we have the conductivity
\begin{equation}\label{eq:sigma}
    \sigma \approx \frac{e^2}{h} \frac{2}{\pi^{1/4}} \frac{n^{7/4} d^{1/2}}{\alpha^{3/2}N}.
\end{equation}
where $d_3 \lesssim d \lesssim d_c $.
At $d \sim d_c \sim n^{-1/2}/\alpha$ or $n \sim (\alpha d)^{-2}$, our conductivity Eq.~\eqref{eq:sigma} 
becomes of the order of
\begin{equation}\label{eq:sigma3}
    \sigma \sim \frac{e^2}{h} \frac{n^{3/2}}{N \alpha^2}
\end{equation}
and with logarithmic accuracy crosses over to the bulk one~\cite{skinner2013a}.
At the charge neutrality point $n = n_p \sim (\alpha N)^{2/3}$, we get the minimum conductivity
\begin{equation}\label{eq:sigma_min}
\sigma_{\min} \sim (e^2/h) (N d^3/\alpha^2)^{1/6}
\end{equation}
which is larger than $e^2/h$ in the range of its validity  $d \gtrsim d_3 = \alpha^{2/3} N^{-1/3}$. 
At $d \sim d_c = \alpha^{-4/3} N^{-1/3}$ our $\sigma_{\min} \sim e^2/h\alpha$ and with logarithmic accuracy crosses over to the bulk one~\cite{skinner2013a}.

It is remarkable that due to Rytova-Chaplik-Entin-Kedysh modification of the Coulomb potential of charged impurity
at large range of electron concentrations $(\alpha N)^{2/3} \lesssim n \lesssim (\alpha d)^{-2} $, not only $\Gamma$, but also the conductivity Eq.~\eqref{eq:sigma}, 
is determined by the long range potential with $q\sim \lambda^{-1} \ll k_F$. Only at large $n$ and $d$ the conductivity Eq.~\eqref{eq:sigma3} is determined the large momentum $q\sim k_F$ scattering on standard Coulomb potentials of impurities located at distances smaller than $k_F^{-1}$ from the TI film surface~\cite{skinner2013a}.

The condition of validity of the above conductivity theory is that the local kinetic energy is larger than $\hbar/\tau$. In the worst case when $\mu=0$ and the local kinetic energy is of the order of $\Gamma(d)$,
using Eqs.~\eqref{eq:gamma_gg1} and ~\eqref{eq:rate3}, we get $\Gamma \tau/\hbar \sim \alpha^{-1/3} (Nd^3)^{1/6} > 1$, because $d > \alpha^{2/3} N^{-1/3}$. This justifies our conductivity results.

\section{Conductivity and hybridization gap}

So far we have ignored the the effect of hybridization gap on the conductivity near neutrality point $\mu=0$. 
In the absence of disorder potential and the Fermi level located inside the hybridization gap TI film becomes an insulator. 
There is a big interest to realize such an insulator in BSTS system~\cite{chong2020,nandi2018}, 
which is expected to show the quantum spin Hall effect and a corresponding four probe resistance $h/2e^2$~\cite{markus2008}.
Below, we suggest a modification of the theory~\cite{nandi2018} of disorder effects on conductivity of the very thin TI film near neutrality point $\mu =0$, using our theory of the random potential developed in Section II. 

We consider the case $\Gamma(d) > \Delta$ so that $\Gamma_{\Delta} \simeq \Gamma(d)$. Then the potential of charged impurities $\phi(\vb{r})$ bends both bands on each surface up and down with characteristic scale $a$, creating at the Fermi level large electron and hole puddles with diameter $\sim a (\Gamma/\Delta)^{4/3}$~\cite{polyakov1995}.
These puddles are separated by thin insulating stripes of the width $x = a\Delta/\Gamma$, which form insulating infinite cluster (see Fig. 4 in Ref.~\cite{nandi2018}) residing at the potential $\phi(\vb{r})$ percolation level $\phi(\vb{r})=0$.

At low temperatures, this system can conduct only if electrons can easily tunnel across these insulating stripes. 
In a  relatively thick TI film where the hybridization gap $\Delta$ is small enough to allow easy tunneling, the conductivity of surface states is still metallic. 
Let us find the upper limit of $\Delta$ for such metallic films, $\Delta_c$. The probability of the Zener-like tunneling across a thin insulating stripe of width $x$ is~\cite{nandi2018}
\begin{align}\label{eq:probability}
P \propto \exp(-\frac{x \Delta}{\hbar v_F}) = \exp(- \frac{a\Delta^2}{\Gamma \hbar v_F}).
\end{align}

For the case $\kappa_f/ \kappa > \alpha^{-1} \gg 1$ studied in Section II using $a=\lambda$, substituting Eqs.~\eqref{eq:gamma_gg1} and \eqref{eq:lambdaneutral} into Eq. \eqref{eq:probability}, we find that the critical value of the hybridization gap at which $P$ loses its exponentially small factor is 
\begin{equation}\label{eq:delta_c2}
\Delta_c(d) = \frac{e^{2}}{\kappa_f d}\frac{(Nd^3)^{1/4}}{ \alpha^{1/2}}.
\end{equation}
Substituting Eq.~\eqref{eq:delta} into Eq.~\eqref{eq:delta_c2} we solved it for the largest 
thickness $d_{\max}$ at which TI films can be considered insulating. 
We got $d_{\max} = 5$ nm for $d_0= 1.2$ nm and $d_{\max} = 8$ nm for $d_0= 1.8$ nm.
In both cases $\Delta_c(d_{\max}) < \Gamma(d_{\max})$, so that our theory is applicable.

For the STO case studied in Section III, $\Gamma=\Gamma_{1}(d)\sim (e^2/\kappa_{f} d) \sqrt{Nd^3}$ and $a=d$. 
Substituting these values into Eq. \eqref{eq:probability}  we arrive to the same $\Delta_c(d)$ Eq.~\eqref{eq:delta_c2}
~\footnote{This universal value is larger than the estimate of Ref.~\cite{nandi2018}, 
where the bulk TI surface screening radius~\cite{skinner2013a} was used for $a$.}.
Thus, our neutrality point conductivity is valid if $d > d_{\max}$, while spin Hall effect
can be achieved only at $d< d_{\max}$. 
This result is in qualitative agreement with results of Ref.~\cite{chong2020}. 
On the other hand, the fact that in Ref.~\cite{nandi2018} the TI film thickness $d=4$ nm was not sufficiently small still remains unexplained.

\begin{acknowledgments}
We are grateful to S.K. Chong, V.V. Deshpande, B. Skinner, and D. Weiss  and  for useful discussions. Y.H. was partially supported by the William I. Fine Theoretical Physics Institute.
\end{acknowledgments}

\medskip

%

\end{document}